\documentstyle[12pt]{article}
\begin{document}

\begin{titlepage}

\begin{flushright} 
EPHOU-96-003

May, 1996

\end{flushright}

\ \vspace {15mm}
\begin{center}
{\Large Quantization of $gl(1,{\bf R})$ Generalized Chern-Simons Theory\\}
{\Large in 1+1 Dimensions\\}
\vspace{1cm}
{\bf {\sc Noboru Kawamoto, Eisaku Ozawa and Kazuhiko Suehiro}}
%                                       \footnote[1]{} 

{\it{ Department of Physics, Hokkaido University }}\\
{\it{ Sapporo, 060, Japan}}

\end{center}
\vspace{2cm}

\begin{abstract}
We present a quantization of previously proposed generalized Chern-Simons 
theory with $gl(1,{\bf R})$ algebra in 1+1 dimensions. 
This simplest model shares the common features of generalized CS theories: 
on-shell reducibility and violations of regularity.
On-shell reducibility of the theory requires us to use 
the Lagrangian Batalin-Vilkovisky 
and/or Hamiltonian Batalin-Fradkin-Vilkovisky formulation. 
Since the regularity condition is violated, their quantization is 
not straightforward. In the present case we can show that both
formulations give an equivalent result. It leads to an interpretation 
that a physical degree of
freedom which does not exist at the classical level appears at the 
quantum level.

\end{abstract}

\end{titlepage}

\renewcommand {\theequation} {\arabic {section}.\arabic{equation}}

%\begin{document}

\setlength{\baselineskip}{7mm}

\section {Introduction} \label {sec:intro}
\setcounter {equation}{0}
%\hspace*{\parindent}

Although three forces in nature are described as gauge theories and 
have been quantized, the quantization of gravity is not yet achieved.
Among the difficulties, nonrenormalizability of Einstein's gravity is
the most difficult one.
There are still some belief that superstring theory describes
real nature containing gravity. 
There is, however, another approach to describe the gravity theory.

Witten has shown that three dimensional gravity can be described
by a Chern-Simons theory with an appropriate algebra which possesses 
a non-degenerate invariant bilinear form (ref.\cite{wit88}).
At the same time he has shown that Chern-Simons theory in three dimensions 
is exactly solvable (ref.\cite{wit89}).
This fact ensures the consistency and finiteness of 
the three dimensional gravity.
Furthermore, it has also been shown that the three dimensional conformal 
gravity can be also treated by the Chern-Simons theory (ref.\cite{howit}).

The extension of the Chern-Simons theory into arbitrary dimensions, 
which was given by one of the authors (N.K.) and Watabiki, was aimed to make 
a similar approach to gravity as in the three dimensional case 
(refs.\cite {kw92-1}-\cite {kw93-1}).
Although their action keeps a formal correspondence with the ordinary 
Chern-Simons theory, it can be used in arbitrary dimensions.
The gauge symmetries are also formally similar to those of ordinary 
Chern-Simons theory, or in other words, the extension is done 
by keeping these similarities, but their explicit expressions are 
very complicated. 
This theory might be useful in formulating four dimensional gravity.
It was shown that topological conformal gravity can be formulated 
by four-dimensional generalized Chern-Simons theory 
(ref.\cite {kw93-1}).
However this theory is generally on-shell reducible and its equations 
of motion violate the regularity in the sense we will explain later.
Thus the quantization of this system is not straightforward.

In this paper we investigate the quantization of the simplest model
defined in two dimensions with $gl(1,{\bf R})$ algebra.
Though this model is quite simple, it shares the common characteristics,
reducibility and violation of the regularity.
The quantization of the simplest model may give a clue for the quantization 
of general cases. We apply methods developed for quantizing reducible systems
with cares for the violation of the regularity.

After the introduction of the model in Sec.~2, we present a quantization
based on a Lagrangian formulation, {\it \`{a} la} Batalin and Vilkovisky 
(refs.\cite {bv81}) in Sec.~3.
Their treatment gives a systematic way to deal with reducible gauge symmetries
and obtain a BRST invariant gauge fixed action by solving so-called
master equation. Even with the violation of the regularity we can utilize 
this formulation for the present model.
However, the meaning of the physical space defined
by the BRST cohomology is not obvious in the Lagrangian formulation.
The meaning of the BRST cohomology and thus the structure of the quantization 
are clearer in the Hamiltonian formalism.
Therefore we investigate the model in the Hamiltonian formulation in Sect.~4.
The systematic treatment in the Hamiltonian formalism is given by
Batalin, Fradkin and Vilkovisky (refs.\cite {fb75}-\cite {bf83-1}),
as an extension of the Dirac's treatment of the constrained systems 
(ref.\cite {dirac}) and a comprehensive explanation can be found
in ref. \cite {henbook}. We use this formulation with a suitable
generalization. We conclude in Sect.~5 with summaries and discussions.

\section { $gl(1,{\bf R})$ model in two dimensions} 
\setcounter {equation}{0}

The generalized Chern-Simons theory which was introduced by N.K.
and Watabiki (refs.\cite {kw92-1}-\cite {kw93-1}) is the generalization of 
the ordinary three dimensional Chern-Simons theory into arbitrary dimensions.
The generalized theory highly respects the structure of the three dimensional 
ordinary Chern-Simons theory.
Indeed, the action of the generalized Chern-Simons theory possesses 
the following form,
\begin {eqnarray}
S = \frac {1}{2} \int _M \ Tr({\cal A}Q{\cal A} + \frac {2}{3} {\cal A}^3),
\end {eqnarray}
where $Q$ corresponds to the ordinary exterior derivative and 
${\cal A}$ to a Lie algebra valued gauge field one form.
However ${\cal A}$ is defined to contain arbitrary forms to extend
the system to arbitrary dimensions. (See refs.\cite {kw92-1}-\cite {kw93-1}
for details.)

The simplest case defined in two dimension is given in ref.\cite {kw92-2}
based on the $gl(1,{\bf R})$ algebra.
The action expanded into components is given by  
\begin {eqnarray}
S = \int \ d^2 x \left( \epsilon ^{\mu \nu} \partial _{\mu} 
        \omega _{\nu} \phi - \frac {1}{2} \epsilon ^{\mu \nu} 
                B _{\mu \nu} \phi ^2 \right), \label {eqn:bi}
\end {eqnarray}
where $\phi$, $\omega_\mu$ and $B_{\mu\nu}$ are scalar, vector and
antisymmetric tensor fields, respectively, 
and $\epsilon^{01}=-\epsilon_{01}=1$ in $1+1$ dimension.
This lagrangian possesses gauge symmetries  
\begin {eqnarray}
\delta _g \phi & = & 0, \label {eqn:bb} \\
\delta _g \omega _{\mu} & = & \partial _{\mu} v + 2 \phi u _{\mu}, \\
\delta _g B & = & \epsilon ^{\mu \nu} \partial _{\mu} u _{\nu}, \label {eqn:bc}
\end {eqnarray}
where $B$ is defined 
by $B \equiv \frac {1}{2} \epsilon ^{\mu \nu} B _{\mu \nu}$.
Here gauge fields and gauge parameters are real variables and thus
the corresponding algebra is $gl(1,{\bf R})$.
Although the action has a simple form, 
it has some unusual properties which are common to generalized CS theories.
The equations of motion of this theory are given by
\begin {eqnarray}
\phi : & & \epsilon ^{\mu \nu} \partial _{\mu} \omega _{\nu} - 2 \phi B = 0, \\
\omega _{\mu} : & & \epsilon ^{\mu \nu} \partial _{\nu} \phi = 0, \\
B : & & - \phi ^2 = 0.
\end {eqnarray}
Generally, equations of motion are called regular 
if any function of field variables vanishing at a stationary point
of the Lagrangian can be written as their ``linear'' combination,
where the coefficients of the combination could be field dependent.
From this definition, it is clear that the regularity is violated 
in the present case.
$\phi$ itself vanishes due to eq.(2.8) while $\phi$ can not 
be written as a ``linear'' combination of the above equations of motion
unless we accept a singular coefficient like ${1\over \phi}$. 

In addition this system is on-shell reducible since the gauge transformations
(2-3)-(2-5) are invariant under the transformation
\begin {eqnarray}
  \delta_\lambda v &=& -2 \phi \lambda,           \\
  \delta_\lambda u_\mu &=& \partial _\mu \lambda,        
\end {eqnarray}
with the on-shell condition $\partial _\mu \phi=0$.
Since the transformations (2.9) and (2.10) are not reducible any more,
the Lagrangian (2.2) is called a first-stage reducible system.

These two aspects, the violation of the regularity and the on-shell 
reducibility, are common features in the generalized Chern-Simons theory.

\section {$gl(1,{\bf R})$ model in the Lagrangian formalism}
\setcounter {equation}{0}
%\hspace*{\parindent}

In this section we present a quantization of the model
based on the Lagrangian formalism.
We can rewrite the action (\ref {eqn:bi}) into the following form
by using $B \equiv \frac {1}{2} \epsilon ^{\mu \nu} B _{\mu \nu}$,
\begin {eqnarray}
S_0 = \epsilon ^{\mu \nu} \partial _{\mu} \omega _{\nu} \phi 
          - B \phi ^2,
\end {eqnarray}
where the integration symbol is omitted. 
The gauge symmetries are given in the previous section, 
eqs.(\ref {eqn:bb})-(\ref {eqn:bc}), 
and this is a first-stage reducible system.

According to the procedure given by  Batalin and Vilkovisky 
(refs.\cite {bv81}), we introduce ghost fields $C$ and $C_\mu$ 
corresponding to the gauge freedom $v$ and $u_\mu$
and a ghost for ghost $C_1$ to the reducible degree $\lambda$, 
which are fermionic and bosonic, respectively.
In addition, a minimal set of anti-fields 
$\Phi_{min}^*=(\phi^*,\omega^{*\mu},B^*,C^*,C_\mu^*,C_1^*)$ 
are introduced corresponding to the field variables
$\Phi_{min}=(\phi,\omega_\mu,B,C,C_\mu,C_1)$.
The Grassmann parities of the anti-fields are opposite to that of
corresponding fields, as usual.

Now the minimal action is obtained by solving the master equation
\begin {eqnarray}
  (S_{min}(\Phi_{min},\Phi_{min}^*),S_{min}(\Phi_{min},\Phi_{min}^*)) &=& 0,
                                        \\
  (X,Y)&=&{\partial_r X \over \partial \Phi^A}
          {\partial_l Y \over \partial \Phi_A^*}
          -{\partial_r X \over \partial \Phi_A^*}
          {\partial_l Y \over \partial \Phi^A},
\end {eqnarray}
with the boundary conditions
\begin {eqnarray}
\left. S_{min}(\Phi_{min},\Phi_{min}^*)\right|_{\Phi_{min}^*=0}&=&S_0,    \\
\left. {\partial_l S_{min}(\Phi_{min},\Phi_{min}^*) \over \partial\phi^*}
         \right|_{\Phi_{min}^*=0}&=&0,     \\
\left. {\partial_l S_{min}(\Phi_{min},\Phi_{min}^*) 
                \over \partial\omega^{\mu *}}
         \right|_{\Phi_{min}^*=0} &=&\partial_\mu C+2\phi C_\mu,     \\
\left. {\partial_l S_{min}(\Phi_{min},\Phi_{min}^*) \over \partial B^*}
         \right|_{\Phi_{min}^*=0} &=& \epsilon^{\mu\nu} \partial_\mu C_\nu,  \\
\left. {\partial_l S_{min}(\Phi_{min},\Phi_{min}^*) \over \partial C^*}
         \right|_{\Phi_{min}^*=0,\, on-shell} &=&0,                   \\
\left. {\partial_l S_{min}(\Phi_{min},\Phi_{min}^*) \over \partial C^{\mu *}}
         \right|_{\Phi_{min}^*=0,\, on-shell} &=& \partial_\mu C_1,
\end {eqnarray}
where $r$ and $l$ denote right and left derivatives, respectively.
It should be noted that the boundary conditions (3.8) and (3.9) are
required only on-shell.\footnote[1]{
      Though it is not manifestly stated in refs.\cite {bv81} that 
      these conditions should be required
      only on-shell, they are enough from 
      the definition of on-shell reducibility.}
It is straightforward to solve the master equation
perturbatively in the order of antighost fields by taking into account 
the fact that $\phi$ equals zero on the shell.
Though it is not assured that the master equation possesses a solution
when the regularity of equations of motion is violated, there is a solution
in the present case. We find
\begin {eqnarray}
S _{min} & = & \epsilon ^{\mu \nu} \partial _{\mu} \omega _{\nu} \phi 
    - B \phi ^2 + \omega ^{* \mu} ( \partial _{\mu} C + 2 \phi C _{\mu} 
    - \epsilon _{\mu \nu} \omega ^{* \nu} C_1), \nonumber \\
& & + B^* \epsilon ^{\mu \nu} \partial _{\mu} C _{\nu} 
    + C ^{* \mu} \partial _{\mu} C_1 - 2 C ^* \phi C_1.
\end {eqnarray}

We next introduce a nonminimal action, which must be added to
the minimal one to make a gauge fixing, as
\begin {eqnarray}
S _{nonmin}  = {\bar C} ^* b + {\bar C} ^*_{\mu} b^{\mu} 
               + {\bar C}_1 ^* b_1 + \eta ^* \pi, 
\end {eqnarray}
where auxiliary fields $b$, $b^\mu$, ${\bar C}_1$ and $\eta$ are bosonic 
and $b_1$, $\pi$, ${\bar C}$ and ${\bar C} _{\mu}$ are fermionic, respectively.
Their corresponding antifields possess the opposite Grassmann parities.
A gauge fixing is done by choosing a suitable gauge fermion \cite {bv81}.
We can adopt, for example, the following gauge fermion $\Psi$, which leads to
a Landau-type gauge fixing:
\begin {eqnarray}
\Psi =  {\bar C} ^{\mu} \partial ^{\nu} B _{\mu \nu} 
       + {\bar C} ^{\mu} \partial _{\mu} \eta 
       + {\bar C}_1 \partial ^{\mu} C _{\mu} 
       + {\bar C} \partial ^{\mu} \omega _{\mu},
\end {eqnarray}
where we assume that the base manifold is flat with the 
metric $\eta_{\mu\nu}=\hbox{diag}(-1,+1)$ for simplicity.
The anti-fields can be eliminated by substituting the derivative 
of the $\Psi$ into them. Then the action turns out to be 
\begin {eqnarray}
S & = & \left.S(\Phi,\Phi^*)\right||_{\Phi^*=
              {\partial \Psi \over \partial \Phi}}        \nonumber \\
  & = &\epsilon ^{\mu \nu} \partial _{\mu} \omega _{\nu} \phi - B \phi ^2 
         - \partial ^{\mu} {\bar C}  (\partial _{\mu} C + 2 \phi C _{\mu}) 
        -  \epsilon _{\mu \nu} \partial ^{\mu} {\bar C} 
                    \partial ^{\nu} {\bar C} C_1  \nonumber \\
    & & + \frac {1}{2} (\partial ^{\mu} {\bar C}^{\nu} 
        - \partial ^{\nu} {\bar C}^{\mu})(\partial _{\mu} C _{\nu} 
        - \partial _{\nu} C _{\mu}) 
        - \partial ^{\mu} {\bar C}_1  \partial _{\mu} C_1   \\
     & &   + b \partial ^{\mu} \omega _{\mu} 
        + b ^{\mu} (\partial _{\mu} \eta 
                 -\epsilon_{\mu \nu} \partial ^{\nu} B )
        - b_1 \partial ^{\mu} C _{\mu} 
        - \partial _{\mu} {\bar C} ^{\mu} \pi.  \nonumber
\end {eqnarray}
It should be noted that the fields ${\bar C}$, ${\bar C}^{\mu}$, $C_1$,
${\bar C}_1$ and $b_1$ are treated as purely imaginary fields to 
make the action hermitian, following to the convention 
of ghost fields $C$,$C_\mu$ being real variables. This action is invariant 
under the following on-shell nilpotent
BRST transformation when applied as a right variation:
\begin {eqnarray}
s \Phi & = & \left. (\Phi, S(\Phi, \Phi^*)) \right|_{\Phi^*=
              {\partial \Psi \over \partial \Phi}}             \\
s \phi & = & 0, \\
s \omega _{\mu} & = & \partial _{\mu} C + 2 \phi C _{\mu} 
       + 2 \epsilon _{\mu \nu} \partial^{\nu} {\bar C} C_1, \\
s B & = & \epsilon ^{\mu \nu} \partial _{\mu} C _{\nu}, \\
s C & = & -2 \phi C_1, \\
s C _{\mu} & = & \partial _{\mu} C_1, \\
s C_1 & = & 0, \\
s {\bar C} & = & b, \qquad\,\,\, s b  = 0,                  \\
s {\bar C}^\mu & = & b^\mu, \qquad s b^\mu =0,      \\
s {\bar C}_1 & = & b_1, \qquad s b_1 = 0,           \\
s \eta & = & \pi, \qquad\,\, s \pi =0.                
\end {eqnarray}

It is easy to see that the action does not receive a quantum correction
and is also a solution of the quantum master equation.
This fact can be naturally understood from a perturbative calculation.
Nonvanishing connected $n$ point functions
for $n\ge 3$ include only three different three-point functions of
unphysical fields, which exist only at the tree level.

%-------------------------------------------------------------------

\section {$gl(1,{\bf R})$ model in the Hamiltonian formalism}
\setcounter {equation}{0}
%\hspace*{\parindent}

The same model as in the previous section is investigated 
by the Hamiltonian formalism in this section.
From the action (3.1) we obtain canonical momenta as 
\begin {eqnarray}
\pi _{\phi} & = & 0, \label {eqn:at} \\
\pi _{\omega _0} & = & 0, \\
\pi _{\omega _1} & = & \phi, \label {eqn:ba} \\
\pi _B & = & 0. \label {eqn:au}
\end {eqnarray}
All of these equations give primary constraints.
The canonical Hamiltonian following from the Lagrangian becomes
\begin {eqnarray}
H _C = \phi \partial _1 \omega _0 + B \phi ^2. \label {eqn:az}
\end {eqnarray}
This Hamiltonian and the above constraints with Lagrange multipliers 
define the following total Hamiltonian,
\begin {eqnarray}
H _T = \phi \partial _1 \omega _0 + B \phi ^2 + \lambda _{\phi} \pi _{\phi} 
        + \lambda _{\omega _0} \pi _{\omega _0} 
        + \lambda _{\omega _1}( \pi _{\omega _1} - \phi) + \lambda _B \pi _B.
\end {eqnarray}
According to the ordinary Dirac's procedure, we have to check the consistency 
of the constraints.
The results are
\begin {eqnarray}
\partial_0 \pi _{\phi} & = & - \partial _1 \omega _0 - 2 \phi B 
               + \lambda _{\omega _1} \ = \ 0, \\
& & \Rightarrow \lambda _{\omega _1} = \partial _1 \omega _0 + 2 \phi B, \\
\partial_0 \pi _{\omega _0} & = & \partial _1 \phi \ = \ 0, \\
\partial_0 (\pi _{\omega _1} - \phi) & = & - \lambda _{\phi} \ = \ 0, \\
& & \Rightarrow \lambda _{\phi} = 0, \\
\partial_0 \pi _B & = & - \phi ^2 \ = \ 0.
\end {eqnarray}
Two lagrange multipliers are determined. 
After the substitution of these expressions into the total Hamiltonian, 
we obtain 
\begin {eqnarray}
H _T = \pi _{\omega _1} \partial _1 \omega _0 + 2 \phi \pi _{\omega _1} B 
- B \phi ^2 + \lambda _{\omega _0} \pi _{\omega _0} + \lambda _B \pi _B.
\end {eqnarray}
At the same time, we have found secondary constraints
\begin {eqnarray}
- \phi ^2 & = & 0. \label {eqn:aw} \\
\partial _1 \phi & = & 0, \label {eqn:av} 
\end {eqnarray}
The consistency check of these constraints gives no other relations.
After all, we have obtained the set of constraints 
eqs.(\ref {eqn:at})-(\ref {eqn:au}) 
and eqs.(\ref {eqn:av}) and (\ref {eqn:aw}).

It should be noted that the constraints (\ref {eqn:av}) and (\ref {eqn:aw})
violates Dirac's regularity condition.
The regularity in Hamiltonian formulation can be stated similarly
to the Lagrangian case by replacing equations of motion by constraint equations.
Constraints are called regular 
if any function of canonical variables vanishing on the constraint surface
can be written as their ``linear'' combination,
where the coefficients of the combination
could be dependent on canonical variables.
In the present case eq.(4.14) implies that $\phi$ vanishes on the constraint
surface. However $\phi$ itself can not be written as a "linear" combination
of (4.14) and (4.15) unless we admit expressions 
like $-{1\over \phi}G_1$ or $\int dx^1 G_2$.
The former expression includes a singular coefficient and the latter
requires to specify a boundary condition. Thus they are not acceptable.

We can replace these two constraints by
a single one
\begin {eqnarray}
\phi=0,
\end {eqnarray}
which is equivalent to eq.(\ref {eqn:aw}) in the present case.
Then  we can separate the constraints into the first class 
and second class. With some redefinitions  
we can obtain a set of constraints as
\begin {eqnarray}
second \ class & & \pi _{\phi} = 0, \ \ \phi = 0, \\
first \ class & & \pi _{\omega _0} = 0, \\
& & \pi _{\omega _1} = 0 , \\
& & \pi _B = 0.
\end {eqnarray}
These constraints imply that no dynamical variables exist, as is expected
from the topological nature of the generalized Chern-Simons action.

Now the quantization of this system is trivial since there is no
dynamical degree of freedom.
(We treat a flat base manifold here and do not take into
account possible finite degrees of freedom depending on the
topology of the base manifold.)
By adopting gauge fixing conditions $\omega_0=\omega_1=B=0$,
all variables and thus Hamiltonian $H_T$ itself vanish identically.
Thus we can conclude this theory is completely empty.

This is, however, not the end of the story. In more general cases with
non-abelian gauge groups, constraints 
of the form $\phi_1^2-\phi_2^2-\phi_3^2=0$ can appear. 
In these cases, we can not adopt constraints which satisfy
the regularity condition and at the same time give a constraint surface 
without singularities.
In the above example, we can linearize the constraint equation 
with respect to $\phi_1$ by choosing 
one of the branches of $\phi_1=\pm \sqrt{\phi_2^2+\phi_3^2}$. It, however, 
gives a singular constraint surface with a conical singularity at $\phi_i=0$.
Therefore it seems rather natural in the generalized Chern-Simons theory
to adopt a quantization method different from the usual one, {\it i.e.},
a quantization based on regularity violating constraints that follow directly
from the Lagrangian. In the following we perform the Hamiltonian BRST
quantization {\it \`{a} la} Batalin, Fradkin and Vilkovisky by using the
regularity violating constraints.
It is interesting that in this treatment of Hamiltonian formulation 
with a suitable choice of gauge condition, we can show that the gauge fixed
action thus obtained is just the same as the result of Lagrangian formulation
in Sect.~3.
Though  the Lagrangian and Hamiltonian constructions
are formally equivalent in usual cases as shown 
in refs.\cite {dejon93-1} and \cite {ad93}, we can show the equivalence
of both formulations when applied to the present model only 
if we adopt the regularity violating constraints in the Hamiltonian formalism.
The physical meaning of the quantization in the Lagrangian formulation
was not obvious in the previous section 
while it becomes clear in the Hamiltonian formulation 
from the above mentioned equivalence.

We first rearrange the constraints (4.1)-(4.4), (4.14) and (4.15) into
(4.1)-(4.4) and
\begin {eqnarray}
G_1 & \equiv & -\pi _{\omega _1}^2  =  0, \label {eqn:ax} \\
G_2 & \equiv &  \partial _1 \pi _{\omega _1}  =  0, \label {eqn:ay}
\end {eqnarray}
so that the first and second class constraints are separated.
Now we can carry on without the variables $\phi$ and $\pi_\phi$ thanks to the
second class constraints (4.1) and (4.3) if we replace 
all $\phi$ by $\pi_{\omega_1}$ and set $\pi_\phi=0$.
We further adopt gauge conditions for the first class constraints (4.2) and 
(4.4) as $\omega_0=B=0$, for simplicity. Though this is not inevitable,
it turns out that this is the simplest way to lead to the gauge fixed
Lagrangian of the form (3.13). Then we can also 
eliminate $\omega_0$, $\pi_{\omega_0}$, $B$ and $\pi_B$ from the system.
After the above manipulation, we have two phase space 
variables $\omega_1$ and $\pi_{\omega_1}$ with the first class constraints
(4.21) and (4.22). Then the total Hamiltonian (4.13) vanishes completely.
It should be noted that these constraints violate the regularity condition
but only in the constant mode of $\pi_{\omega_1}$. 

Now following to the procedure in the Hamiltonian formalism,
we define a Koszul-Tate differential $\delta$ and Grassmann odd 
ghost momenta $P_1$ and $P_2$ as 
\begin {eqnarray}
\delta \omega _1 &=& \delta \pi_{\omega _1}=0,      \\
\delta P _1 & = & - \pi _{\omega _1}^2, \\
\delta P _2 & = & \partial _1 \pi _{\omega _1}.
\end {eqnarray}
In the case the regularity condition is not violated, Koszul-Tate 
differential is defined so that its homology is a set of functions 
on the constraint surface, {\it i.e.} those not vanishing on the surface.
We follow the usual procedure and give a comment afterwards 
on what the violation of regularity causes.
Constraints (4.21) and (4.22) are not independent 
due to the following relation 
\begin {eqnarray}
\partial_1 G_1 + 2 \pi_{\omega_1} G_2=0,
\end {eqnarray}
which shows that they are reducible. 
In the present case there is only one independent relation 
and thus the system is called first stage reducible. 
Then it is necessary to introduce one more ghost momentum $P$, 
which is Grassmann even, as
\begin {eqnarray}
\delta P = -  ( \partial _1 P _1 + 2 \pi _{\omega _1} P _2 ).
\end {eqnarray}
This differential is nilpotent by definition.
%and antighost numbers are assigned as 0,0,1,1 and 2 for 
%$\omega _1$, $\pi_{\omega _1}$, $P_1$, $P_2$ and $P$, respectively.
Note that due to the violation of the regularity condition 
the $\delta$-closed constant mode of $\pi_{\omega_1}$ is not $\delta$ exact
while ``classical'' constraints  (4.21) 
requires $\pi_{\omega_1}=0$ on the surface.
This is an interesting phenomenon that a degree of freedom
which does not exist at the ``classical'' level appears 
at the ``quantum'' level.

We next define an extended longitudinal differential $D$ in the following way. 
First the longitudinal differential $d$ is defined from the 
form of the gauge transformations as
\begin {eqnarray}
d \omega _1 & = & \partial _1 \eta _2+ 2 \pi_{\omega_1} \eta_1, \\
d (others) & = & 0.
\end {eqnarray}
where we introduce two fermionic ghost fields of ghost number 1
corresponding to the first class constraints which are 
generators of the gauge transformation.
Due to the reducibility (4.26), ghost fields appear only in the combination
of the right hand side of eq.(4.28). Thus there exists a cohomology of $d$ 
in the ghost number 1 sector of the ghost combination orthogonal 
to the above one. In order to kill this cohomology,
we have to introduce an auxiliary differential 
$\Delta$ and one bosonic ghost $\eta$ 
\begin {eqnarray}
\Delta \eta _1 & = &  \partial _1 \eta, \\
\Delta \eta _2 & = & -2 \pi_{\omega_1} \eta, \\
\Delta (others) & = & 0.
\end {eqnarray}
Now three ghosts $\eta_1$, $\eta_2$ and $\eta$ are introduced.
This makes possible to extend the phase space to include $P$'s and $\eta$'s.
Then $D$ is obtained by the sum of these two differentials,
\begin {eqnarray}
D & = & \Delta + d.
\end {eqnarray}
This differential $D$ is nilpotent in the space of ghosts and 
constraint surface, in other words, nilpotent modulo $\delta$-exact term.

Finally the BRST differential $s$ is defined by $\delta$ and $D$,
\begin {eqnarray}
s & = & \delta + D + \stackrel {(1)}{s}, \\ 
& & \delta \eta_1 =\delta \eta_2 = \delta \eta= 0, \\
& & D P_1 = D P_2 = D P = 0.
\end {eqnarray}
The $\stackrel {(1)}{s}$ is determined by the requirement 
of the nilpotency of $s$.
Since $D$ is nilpotent modulo $\delta$-exact term, we can realize
the nilpotency of $s$ by choosing
\begin {eqnarray}
\stackrel {(1)}{s} \omega _1 & = & 2 \eta P _2, \\
\stackrel {(1)}{s} (others) & = & 0.
\end {eqnarray}
The action of the differential $s$ is listed as follows:
\begin {eqnarray}
s P _1 & = & - \pi _{\omega _1}^2, \\
s P _2 & = & \partial _1 \pi _{\omega _1}, \\
s P & = & - (\partial _1 P _1 + 2 \pi _{\omega _1} P _2), \\
s \omega _1 & = & \partial _1 \eta _2 + 2 \pi _{\omega _1} \eta _1 
                     + 2 \eta P _2, \\
s \pi_{\omega_1} & = & 0,    \\
s \eta _1 & = &  \partial _1 \eta, \\
s \eta _2 & = & -2 \pi _{\omega _1} \eta, \\
s \eta & = & 0.
\end {eqnarray}
The extended phase space is defined to include 
the above ghosts and ghost momenta with a canonical structure
\begin {eqnarray}
[P_i, \eta _j]= - \delta _{ij}, \ \ [P, \eta]=-1,
\end {eqnarray}
where $[,]$ represents the graded Poisson bracket which will be
replaced by the graded commutation relation multiplied by $-i$ as 
in usual cases.
By using this canonical relation, 
the nilpotent BRST charge $\Omega ^{Min}$ is defined by 
\begin {eqnarray}
\Omega ^{Min} & = & \eta _1 \pi _{\omega _1}^2 
                  - \eta _2 \partial _1 \pi _{\omega _1} 
                  + \eta (\partial _1 P _1 + 2 \pi _{\omega _1} P _2),
\end {eqnarray}
which realizes $s X = [ X, \Omega^{Min}]$.

In order to fix the gauge, 
we have to extend the phase space further and introduce 
the following set of canonical variables and their momenta,
%$\lambda_a$, $b_a$, $\bar C_a$ and $\rho_a$ with $a=0,1,2,3$.
%
\begin {eqnarray*}
\lambda _1,\lambda _2, \ \ b _1,b _2, & & {\bar C} _1,{\bar C} _2, 
                \ \ \rho_1,\rho_2, \\ 
\lambda _0, \ \ \ \ b _0, \ & & \ \ {\bar C} _0, \ \ \ \ \rho_0, \\
\lambda, \ \ \ \ b, \ & & \ \ {\bar C}, \ \ \ \ \rho. 
\end {eqnarray*}
Two sets with indices 1 and 2 correspond to two first class constraints and
the other {\it two} sets to {\it one} reducible condition (4.26).
Their statistics are bosonic for $b_i$, $\lambda_i$, $\bar C$ and $\rho$, and
fermionic for ${\bar C}_i$, $\rho_i$, $b$ and $\lambda$.
The canonical structure is defined by
\begin {eqnarray}
[ \rho_i , {\bar C }_j ] & = & [b_i, \lambda_i] = - \delta _{ij},  \\ [0pt]
[ \rho , {\bar C} ] & = & [b, \lambda] = - 1.
\end {eqnarray}
The action of BRST differential is also extended to these variables as
\begin {eqnarray}
s \lambda_a & = & -\rho_a, \ \ s \rho_a = 0 \\
s {\bar C}_a & = & b_a, \ \ s b_a=0,
\end {eqnarray}
where $a$ denotes $i=0,1,2$ or {\it nothing}. 
The corresponding extended BRST charge is given by
\begin {eqnarray}
\Omega & = & \Omega ^{Min} + \Omega ^{Nonmin}, \\
\Omega ^{Nonmin} & = & - \sum_{i=0}^2  \rho_i b_i+ \rho b.
\end {eqnarray}

Now the gauge fixed action $S$ is obtained by a Legendre transformation
from the Hamiltonian in the extended phase space:
\begin {eqnarray}
S & = & {\dot \omega} _1 \pi _{\omega _1} 
       + \sum_{i=1,2} {\dot \eta} _i P _i + {\dot \eta} P 
       + \sum_{i=0}^2 ( {\dot \lambda}_i b _i + {\dot {\bar C}}_i \rho_i )
       + {\dot \lambda} b + {\dot {\bar C}} \rho - H_K,    \\
H_K & = & [K,\Omega],
\end {eqnarray}
where $K$ is called a gauge-fixing fermion. The gauge fixed 
Hamiltonian $H_K$ consists of gauge-fixing and ghost parts only since
the total Hamiltonian of the system have vanished.
The gauge-fixing fermion generally takes 
the form $K=\bar C_a \chi_a-P_i\lambda_i$ with $\chi_a$ being
functions of phase space variables corresponding to a gauge choice. 
(In the present case, there are four terms in the first sum 
and three in the second.)
There is no systematic way to find $K$ so as to yield a covariant
expression. Here, however, we can use the result in the Lagrangian
formulation as a clue. Actually we want to show that the two formulations
give an equivalent result. Therefore we plug the above form into (4.55)
and compare the form with the expression (3.13). 
We have found that the following gauge-fixing fermion $K$ works as desired:
\begin {eqnarray}
K = -{\bar C} _1 \partial _1 \lambda _0  + {\bar C} _2 \partial _1 \omega _1 
    - {\bar C} _0 \partial _1 \lambda _1 - {\bar C} \partial _1 \eta _1 
    - P_1 \lambda _1 - P_2 \lambda _2 - P \lambda.
\end {eqnarray}
Indeed after integrating out the momentum 
variables $P _1,P _2,P,\rho_1,\rho_2,\rho$ with this gauge-fixing fermion, 
the action becomes 
\begin {eqnarray}
S & = & {\dot \omega} _1 \pi _{\omega _1} 
    + \sum_{i=0}^3 {\dot \lambda} _i b _i + {\dot \lambda} b 
    +  {\dot {\bar C}} _1 (-{\dot \eta} _1 + \partial_1 \lambda)
    -  {\dot {\bar C}} _2 ({\dot \eta} _2 +2 \eta \partial _1 {\bar C} _2 
        - 2 \pi _{\omega _1} \lambda) + {\dot {\bar C}} _0 \rho_0 \nonumber \\ 
& & - {\dot {\bar C}}{\dot \eta} 
    -{\bar C} _1 \partial _1 \rho _0
    - {\bar C} _2 \partial _1 (\partial _1 \eta _2 
            + 2 \pi _{\omega _1} \eta _1) 
    + {\bar C} _0 \partial _1 ({\dot \eta} _1 - \partial_1 \lambda)
    - {\bar C} \partial _1 \partial _1 \eta  \nonumber \\
& & + b _1 \partial _1 \lambda _0 
    - b _2 \partial _1 \omega _1  
    + b _0 \partial _1 \lambda _1 
    +  b \partial _1 \eta _1 
    - \lambda _1 \pi _{\omega _1}^2
    + \lambda _2 \partial _1 \pi _{\omega _1}. 
\end {eqnarray}
If we rename the variables as,
\begin {eqnarray}
\pi _{\omega _1} & \rightarrow & \phi, \\
(\lambda _1, \lambda _2, \lambda_0, \lambda ) & \rightarrow & 
         (B, \omega _0, \eta, C_{\mu=0}), \\
(b _1, b_2, b_0, b) &  \rightarrow & (b^{\mu=1}, -b, b^{\mu=0}, -b_1), \\
(\eta _1, \eta_2, \eta) & \rightarrow & (C_{\mu=1},C,C_1), \\
({\bar C} _1,{\bar C} _2,{\bar C} _0,{\bar C}) & \rightarrow & 
        ({\bar C}^{\mu=1} , - {\bar C},{\bar C}^{\mu=0}, -{\bar C}_1 ),\\
\rho_0 & \rightarrow & -\pi,
\end {eqnarray}
this action  completely coincides with the gauge fixed action (3.13)
in the Lagrangian formalism. Note that we have not taken care of 
the fact that  ${\bar C}$, ${\bar C}^{\mu}$, $C_1$, ${\bar C}_1$ \
and $b_1$ should be treated as purely imaginary fields. The equivalence,
however, continues to hold even if we insert some $i$'s to take it into
account. 

The physical space is defined by a cohomology of BRST differential.
As we have mentioned above, there is a constant mode 
of $\phi (=\pi_{\omega_1})$ which belongs to the homology of the
Koszul-Tate differential $\delta$. At the same time this constant mode
belongs to the cohomology of the differential $D$. Therefore this constant
mode linear in $\phi$ is a physical degree of freedom
while a constant mode of $\phi^2$ is not.
This implies that a physical degree of freedom
which does not exist at the classical level appears at the quantum level.

\section { Conclusions} 
\setcounter {equation}{0}

We have investigated the quantization of the generalized Chern-Simons theory
in the simplest model with $gl(1,{\bf R})$ algebra in 1+1 dimensions
both in the Lagrangian and Hamiltonian formulations. 
Since this type of theory always violates the regularity condition, 
its quantization would be rather naturally defined
by adopting regularity violating equations. 
Thus in this simplest model the square of the zero form field $\phi$, 
which appears as the multiplier of the highest form field and, 
at first sight, looks singular, is treated as one of reducible functions.
We have found that the BRST invariant gauge-fixed action 
obtained from the Lagrangian BV formulation coincides with that from the
BFV Hamiltonian formulation 
when we adopt these regularity violating constraints.
It is surprising that a physical degree of freedom which does not exist
at the classical level appears at the quantum level due to the violation
of the regularity. 
We know that a zero form field plays an important role 
in more realistic generalized CS theories as emphasized 
in the classical discussions (refs.\cite {kw92-1}-\cite {kw93-1}).
In particular the constant part of the zero form field 
became a physical order parameter of the generalized CS theory.
There might be some connections between these two facts.
It is thus important to investigate more realistic cases with non-abelian
gauge algebra and/or in higher space-time dimensions to see a physical
implication of this type of quantization.

We have taken into account neither the metric dependence, 
which appear in the gauge fixing part, nor the dependence on the global 
topology. These points will also be discussed in the future publication.

\vskip 1cm

\noindent{\Large{\bf Acknowledgments}}
%\hspace*{\parindent}

We would like to thank Y. Watabiki and H. Niwamoto for instructive discussions
at the early stage of this investigation.

\end{document}